\documentclass[prl,twocolumn,letterpaper,10pt,twoside,tightenlines,nofootinbib,showpacs]{revtex4}
\usepackage{amsmath,amsfonts,amssymb,latexsym}
\usepackage{graphicx,color}
\usepackage[sort&compress]{natbib}

\begin{document}

\newcommand{\eg}{{\it e.g.}}
\newcommand{\cf}{{\it cf.}}
\newcommand{\etal}{{\it et. al.}}
\newcommand{\ie}{{\it i.e.}}
\newcommand{\be}{\begin{equation}}
\newcommand{\ee}{\end{equation}}
\newcommand{\bea}{\begin{eqnarray}}
\newcommand{\eea}{\end{eqnarray}}
\newcommand{\bef}{\begin{figure}}
\newcommand{\eef}{\end{figure}}
\newcommand{\bce}{\begin{center}}
\newcommand{\ece}{\end{center}}
\newcommand{\red}[1]{\textcolor{red}{#1}}

\newcommand{\dd}{\text{d}}
\newcommand{\ii}{\text{i}}
\newcommand{\lsim}{\lesssim}
\newcommand{\gsim}{\gtrsim}
\newcommand{\RAA}{R_{\rm AA}}

\title{Bottom hadro-chemistry in high-energy hadronic collisions}

\author{Min~He$^{1}$ and Ralf~Rapp$^2$}
\affiliation{$^1$Department of Applied Physics, Nanjing University of Science and Technology, Nanjing~210094, China}
\affiliation{$^2$Cyclotron Institute and Department of Physics and
Astronomy, Texas A\&M University, College Station, Texas 77843-3366, U.S.A.}

\date{\today}

\begin{abstract}
The hadro-chemistry of bottom quarks ($b$) produced in hadronic collisions encodes valuable information on the mechanism of color-neutralization in these reactions. Since the $b$-quark mass is much larger than the typical hadronic scale of $\sim$1\,GeV, $b\bar b$ pair production is expected to be well separated from subsequent hadronization processes. A significantly larger fraction of $b$ baryons has been observed in proton-proton ($pp$) and proton-antiproton ($p\bar{p}$) reactions relative to $e^+e^-$ collisions, challenging theoretical descriptions. We address this problem by employing a statistical hadronization approach with an augmented set of $b$-hadron states beyond currently measured ones, 
guided by the relativistic quark model and lattice-QCD computations. Assuming {\it relative} chemical
equilibrium between different $b$-hadron yields, thermal densities are used as fragmentation weights of $b$-quarks
into various hadron species. With quark model estimates of the decay patterns of excited states, the fragmentation fractions of
weakly-decaying $b$ hadrons are computed and found to agree with measurements in $p\bar{p}$ collisions at the Tevatron. By combining
transverse-momentum ($p_T$) distributions of $b$-quarks from perturbative QCD with thermal 
weights and independent fragmentation toward high $p_T$, a fair description of the $p_T$-dependent $\bar{B}_s^0/B^-$ and $\Lambda_b^0/B^-$ ratios measured in $pp$ collisions at the LHC is obtained. The observed enhancement of $\Lambda_b^0$ production is attributed to the feeddown from thus far unobserved excited $b$ baryons. Finally,  we implement the hadro-chemistry into a strongly-coupled transport approach for $b$-quarks in
heavy-ion collisions, utilizing previously determined $b$-quark transport coefficients in the Quark-Gluon Plasma, to highlight the modifications of
hadro-chemistry and collective behavior of $b$ hadrons in Pb-Pb collisions at the LHC.
\end{abstract}


\maketitle

{\it Introduction.---}
The masses of charm ($c$) and especially bottom ($b$) quarks are much greater than the nonperturbative scale of Quantum Chromodynamics (QCD),
$\Lambda_{\rm QCD}$,  and therefore their production in experiment offers valuable tests of perturbative-QCD
dynamics~\cite{Norrbin:2000zc,Mangano:1991jk}. However, the heavy-quark (HQ) conversion into heavy-flavor (HF) hadrons is an intrinsically soft
process that usually requires phenomenological modeling of nonperturbative fragmentation functions (FFs) to describe the production yields and
momentum spectra of the observed hadrons~\cite{Norrbin:2000zc,Kartvelishvili:1977pi,Peterson:1982ak,Braaten:1994bz}. Fragmentation fractions of
heavy quarks into weakly-decaying heavy hadrons, which include feeddown from excited states via strong or electromagnetic decays, provide a
critical test of hadronization mechanisms, and are commonly denoted as
$f_u$, $f_d$, $f_s$ and $f_{\rm baryon}$, representing the probabilities of, \eg, a $b$ quark hadronizing into a $B^-$, $\bar{B}^0$, $\bar{B}_s^0$
meson and a $b$ baryon (or their charge-conjugate counterparts), respectively. Precise knowledge of these fractions is also important to improve the
sensitivity of searches for physics beyond the standard model via rare decays of $b$ hadrons~\cite{CMS:2014xfa}.

Traditionally, $b$-quark fragmentation has been assumed to be universal across different colliding systems based on the notion that hadronization occurs nonperturbatively at the scale of $\Lambda_{\rm QCD}$~\cite{Lisovyi:2015uqa} independent of the environment. This is supported, within uncertainties, by
measurements of the $f_s/f_d$ ratio that are consistent between $e^+e^-$ collisions at the $Z^0$ resonance at LEP~\cite{ALEPH:1995npv,DELPHI:2003pao}
and $pp$ collisions at the LHC~\cite{LHCb:2011ldp,LHCb:2011leg,LHCb:2013vfg,ATLAS:2015esn,LHCb:2019fns,LHCb:2019lsv,LHCb:2021qbv,ALICE:2021mgk,CMS:2022wkk}. However, a substantially larger value of $f_{\Lambda_b}/f_d$ has been observed in high-energy $b$ jets produced in
$p\bar{p}$~\cite{CDF:2008yux} and $pp$~\cite{LHCb:2014ofc,LHCb:2015qvk,LHCb:2019fns}, compared to $b$ jets from $Z^0$
decays~\cite{ALEPH:1997jlh,DELPHI:2003pao,HFLAV:2019otj}, thus challenging the universality assumption. Similar discrepancies have been
reported in the charm ($c$) sector~\cite{ALICE:2017thy,ALICE:2020wfu,ALICE:2021dhb}.

In practice, FFs for $b$ or $c$ hadrons are usually inferred from $e^+e^-$ annihilation data. Employing these FFs gives a satisfactory
description of $p_T$-differential cross sections for $b$ and $c$ mesons in hadronic collisions within various calculational schemes, such as
Fixed Order Next-to-Leading Logarithm (FONLL)~\cite{Frixione:2007nw,Cacciari:2012ny,Catani:2020kkl}, $k_T$-factorization~\cite{Catani:1990eg,Collins:1991ty,Maciula:2018iuh} or general-mass variable-flavor number scheme (GM-VFNS)~\cite{Kniehl:2005mk,Kniehl:2011bk,Kniehl:2020szu}.
However, the application of FFs for $c$ baryons substantially underestimates $\Lambda_c$ production, especially at low $p_T$~\cite{Maciula:2018iuh,Kniehl:2020szu}, in $pp$ collisions at the LHC~\cite{ALICE:2017thy,ALICE:2020wfu}, further questioning their universality. An extraction of the FFs for $b$ baryons is currently lacking~\cite{Kramer:2018rgb}.

Effects of the partonic environment on HQ hadronization have first been put forward in elementary hadronic collisions~\cite{Hwa:1994uha,Cuautle:1997ti, Braaten:2002yt,Rapp:2003wn,Berezhnoy:2013qln,Christiansen:2015yqa,He:2019vgs}. Specifically, HQ hadronization
may be affected through recombination with valence quarks in the initial state~\cite{Hwa:1994uha,Cuautle:1997ti, Braaten:2002yt,Rapp:2003wn,Berezhnoy:2013qln}, or multi-parton interactions in the final state~\cite{Christiansen:2015yqa}.
In $pp$ collisions at LHC energies, this has been pursued via a statistical coalescence production of $c$ hadrons~\cite{Andronic:2009sv,He:2019tik}, where
hadron yields are determined by the thermo-statistical weights governed by their masses at a universal hadronization
``temperature"~\cite{Braun-Munzinger:2003pwq,Becattini:2009sc,Andronic:2017pug}.
In the present work, we generalize this approach to the bottom sector to compute the $b$ hadro-chemistry, using a large set of $b$-hadron states
that goes well beyond the currently observed spectrum~\cite{ParticleDataGroup:2020ssz}.
By further employing quark model estimates of the decay systematics of excited $b$ hadrons, we are able to predict a large set of fragmentation fractions of weakly-decaying $b$ hadrons.
We also evaluate the  $p_T$ dependence of the hadro-chemistry via a combined recombination/fragementation scheme, which enables
predictions for the total $b\bar{b}$ cross section as well as for the $\bar{B}_s^0/B^-$ and $\Lambda_b^0/B^-$ ratios in $pp$ collisions.
Finally, we implement the new hadro-chemistry into our Langevin transport approach for heavy-ion collisions and highlight predictions for the nuclear modification factor of selected $b$ hadrons in 5\,TeV Pb-Pb collisions.

{\it Bottom-hadron spectrum and strong decays.---}
The experimental effort to search for missing resonances in the HF sector has been ongoing for
decades~\cite{Crede:2013kia,Chen:2016spr,Charles:2016rim}. The current particle data group (PDG) listings are rather
scarce especially for $b$ baryons~\cite{ParticleDataGroup:2020ssz}. Many additional $b$ hadrons are predicted by quark model
studies~\cite{Ebert:2009ua,Ebert:2011kk,Roberts:2007ni} and in good agreement with lattice-QCD (lQCD) results~\cite{Lewis:2008fu,Brown:2014ena}.
We therefore employ a statistical hadronization model (SHM) using two different sets of $b$-hadrons as input: (a) PDG-only states~\cite{ParticleDataGroup:2020ssz} and (b) a relativistic quark model (RQM)~\cite{Ebert:2009ua,Ebert:2011kk} which additionally includes 18 $B$'s, 16 $B_s$'s, 27 $\Lambda_b$'s, 45 $\Sigma_b$'s, 71 $\Xi_b$'s, and 41 $\Omega_b$'s, 
up to meson (baryon) masses of 6.5 (7)\,GeV.
Since we are mostly concerned with the {\it relative} production yields of $b$ hadrons, we use the grand-canonical version of SHM, which works well
for bulk hadron production in minimum-bias $pp$ collisions at the LHC energies~\cite{Sharma:2018jqf,Das:2016muc} (the smallness of the total number
of $b$ hadrons requires a canonical treatment of the $b$ number when computing the {\it absolute} yields, but the induced canonical suppression factor is common to all $b$ hadron containing a single $b$ quark and thus cancels out in hadron ratios~\cite{Chen:2020drg}; likewise, the $b$ fugacity factor, which is  fixed by the total $b\bar{b}$ cross section, is dropped). The thermal density of a given $b$ hadron of mass $m_i$ and spin-isospin degeneracy $d_i$ and containing $N_s^i$ strange or antistrange quarks is then evaluated at the hadronization temperature $T_H$ as
\begin{equation}
n_i^{\rm primary}=\frac{d_i}{2\pi^2}\gamma_s^{N_s^i}m_i^2T_{H}K_2(\frac{m_i}{T_H}) \ ,
\label{primary_density}
\end{equation}
where $K_2$ is the modified Bessel function of second kind and $\gamma_s$$\sim$0.6~\cite{STAR:2008med,He:2019tik} the strangeness suppression
factor in elementary reactions. While the SHM analysis of light-hadron yields in heavy-ion collisions at the LHC~\cite{Andronic:2017pug} indicates a hadronization temperature very comparable to the pseudo-critical chiral transition temperature $T_{\rm pc}^{\chi}$$\sim$155\,MeV determined in lQCD~\cite{HotQCD:2014kol,Borsanyi:2013bia}, a higher hadronization temperature $T_H$$\sim$170\,MeV appears to be more appropriate for HF hadrons in elementary
reactions~\cite{Andronic:2009sv}. A flavor hierarchy in the effective hadronization temperature has also been suggested based on lattice calculations
of quark flavor susceptibilities~\cite{Bellwied:2013cta}.
In the following, we therefore use $T_H$=170\,MeV as the default value and $T_H$=160\,MeV as part of our error estimate.

\begin{figure} [!t]
\includegraphics[width=0.99\columnwidth]{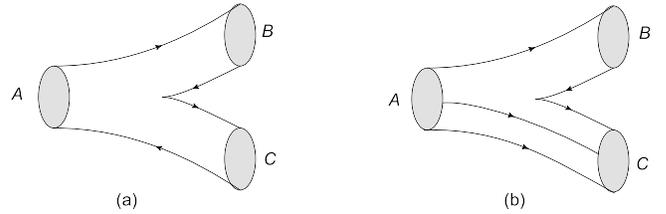}
\vspace{-0.15cm}
\caption{The (a) meson and (b) baryon decay process $A\rightarrow B+C$ in the $^3P_0$ model.}
\label{fig_P03model}
\end{figure}
%
%
%
%
Information about decays of excited $b$ hadrons into their weakly-decaying ground states is very limited, even for observed
states~\cite{ParticleDataGroup:2020ssz}. Instead of taking rather incomplete results available in the literature, we estimate the branching ratios ($BR$s)
of the OZI-allowed strong decays for all employed $b$ hadrons within the $^3P_0$ pair creation model~\cite{Capstick:2000qj}. As schematically shown in Fig.~\ref{fig_P03model}, a pair of quarks with $J^{PC}=0^{++}$ is created from the vacuum and regroups with the quarks within the initial $b$ hadron
into the outgoing meson or baryon. For example, in the case of a baryon decay, there are three ways of quark regrouping~\cite{Chen:2007xf}, leading to a
$b$-baryon plus a light meson or a $b$-meson plus a light baryon in the final state, cf.~Fig.~\ref{fig_P03model}(b). We do not attempt to perform full calculations using realistic hadron wavefunctions~\cite{Roberts:1992esl,Ferretti:2015rsa,Yu:2022ymb}, but obtain the needed $BR$s via counting all possible diagrams of
the types in Fig.~\ref{fig_P03model} once a decay channel opens up, by assuming the number of diagrams for the channel to be proportional to its $BR$.
The probability of creating a quark pair is assumed to be $\propto e^{-2m_q/T_H}$; therefore a diagram involving creating a $s\bar{s}$ is weighted
by $e^{-2m_s/T_H}/e^{-2m_{u/d}/T_H}\simeq1/3$ taking the current-quark masses $m_{u/d}$$\simeq$8\,MeV and $m_s$$\simeq$100\,MeV (this estimate is robust even if constituent-quark masses are used).
We account for direct three-body decays of $b$-baryons by counting the regroupings resulting from creation of two $u\bar{u}$ or $d\bar{d}$ pairs and find significant $BR$s for the
$\Lambda_b^0 + n\pi$ channels of the excited $\Lambda_b$'s or $\Sigma_b$'s decays, which is supported by the measured information on
$\Lambda_b(5912)^0$ and $\Lambda_b(5920)^0$ (reminiscent of $\Lambda_c(2595)^+$ and $\Lambda_c(2625)^+$ in the $c$ sector)~\cite{ParticleDataGroup:2020ssz}. While low-lying excited states all end up in their corresponding ground states (including the pure
electromagnetic decay of $B^*$ and $B_s^*$), higher states have significant cross-feeddowns, \eg, $BR\sim14$\% for excited $B$'s to $\bar{B}_s^0+K$,
$\sim$85-100\% for excited $B_s$'s to $B^-$$/$$\bar{B}^0+K$, $\sim$20-30\% for excited $\Lambda_b$'s ($\Sigma_b$'s) to $B^-(\bar{B}^0)+N$,
$\sim$20\% ($20$\%) for excited $\Xi_b$'s to $\Lambda_b^0+\pi+K$ ($B^-$$/$$\bar{B}^0+\Sigma$), and $\sim75$\% ($\sim25$\%) for excited
$\Omega_b$'s to $\Xi_b^{0-}+K$ ($B^-$$/$$\bar{B}^0+\Xi$). These estimates are overall consistent with available results from a relativized
quark model~\cite{Ferretti:2015rsa,Yu:2022ymb}.


{\it Bottom-hadron fractions and ratios.---}
With the $BR$s as estimated above, the total densities of the weakly-decaying ground states are obtained via
\begin{equation}
n_{\alpha}=n_{\alpha}^{\rm primary} + \sum_in_i^{\rm primary}\cdot BR(i\rightarrow\alpha) \ .
\label{total_density}
\end{equation}
These densities are converted into fractions of the total $b$ content in Table~\ref{tab_frac}, under the constraint of $f_u+f_d+f_s+f_{\Lambda_b^0}+f_{\Xi_b^{0,-}}+f_{\Omega_b^-}=1$ when neglecting the tiny fractions of states made of two or more heavy quarks (\eg, $B_c$ mesons, doubly-bottom baryons or bottomonia). When going from the PDG to the RQM scenario, a marked transfer of the $b$ content from the meson to the baryon sector occurs,
comparable to the experimentally observed $b$-hadron fractions in $e^+e^-$ vs. $p{\bar p}$ collisions~\cite{HFLAV:2019otj} (and reminiscent of the
charm sector~\cite{ALICE:2021dhb}). Specifically, the fraction of $B^-$ ($\bar{B}_s^0$) is reduced by $\sim$10(15)\%, but the fractions of
$\Lambda_b^0$ and $\Xi_b^{0,-}$ are both enhanced by $\sim$50\% upon inclusion of additional baryons in the RQM at $T_H$=170\,MeV, relative
to the PDG scenario (similarly at $T_H$=160\,MeV). The weakly-decaying $b$-hadron fractions obtained in the RQM for both $T_H$=170 and $160$\,MeV turn out to agree with the measurements in $p\bar{p}$ collisions at the Tevatron within uncertainties, $f_u=f_d=0.340\pm0.021$, $f_s=0.101\pm0.015$ and $f_{\rm baryon}=0.220\pm0.048$~\cite{HFLAV:2019otj}.

The calculated ratios of $\bar{B}^0$, $\bar{B}_s^0$, $\Lambda_b^0$ and $\Xi_b^{0,-}$ to $B^-$ are summarized in Table~\ref{tab_rat}.
The mesonic ratios are rather stable, but baryonic ratios are more sensitive to variations in the hadronization temperature. While an equal
production of $B^-$ and $\bar{B}^0$ always holds due to isospin symmetry, the $\bar{B}_s^0/B^-$ ratio is reduced by $\sim$7\% upon inclusion
of additional states in the RQM scenario. The most pronounced effect is caused by the inclusion of missing baryons, enhancing the baryonic ratios by
$\sim$60\% relative to the PDG scenario, leading to $\Lambda_b^0/B^-\sim0.51$ with $T_H$=170\,MeV, rather comparable to $f_{\Lambda_b^0}/(f_u+f_d)=0.259\pm0.018$ as measured by LHCb in 13\,TeV $pp$ collisions~\cite{LHCb:2019fns}.

\begin{table}[!t]
\begin{tabular}{lcccccc}
\hline\noalign{\smallskip}
$f_{\alpha}$        &$B^-$      &$\bar{B}^0$     &$\bar{B}_s^{0}$    &$\Lambda_b^0$     &$\Xi_b^{0,-}$  \\
\noalign{\smallskip}\hline\noalign{\smallskip}
PDG(170)     & 0.3697    &   0.3695    &  0.1073     &  0.1157    &   0.03698      \\
PDG(160)     & 0.3782    &   0.3780    &  0.1094     &  0.1023    &   0.03144      \\
RQM(170)     & 0.3391    &   0.3389    &  0.09152    &  0.1737    &   0.05503      \\
RQM(160)     & 0.3533    &   0.3532    &  0.09620    &  0.1502    &   0.04565      \\
\noalign{\smallskip}\hline
\end{tabular}
\caption{Fractions of ground-state $b$ hadrons (relative to total $b\bar b$) from the SHM with $T_{H}$=170 and 160\,MeV in the
PDG and RQM scenarios. $\Xi_b^{0,-}$ denotes the sum of two isospin states. The $\Omega_b^-$ fraction is  $\sim$0.1\% (not shown here).}
\label{tab_frac}
\end{table}
\begin{table}[!t]
\begin{tabular}{lcccccc}
\hline\noalign{\smallskip}
$r_{\alpha}$        &$\bar{B}^0/B^-$    &$\bar{B}_s^{0}/B^-$    &$\Lambda_b^0/B^-$   &$\Xi_b^{0,-}/B^-$ \\
\noalign{\smallskip}\hline\noalign{\smallskip}
PDG(170)      & 0.9995    & 0.2904     &  0.3129   &  0.1000    \\
PDG(160)      & 0.9995    & 0.2894     &  0.2706   &  0.08313    \\
RQM(170)      & 0.9994    & 0.2699     &  0.5122   &  0.1623     \\
RQM(160)      & 0.9996    & 0.2723     &  0.4250   &  0.1292     \\

\noalign{\smallskip}\hline
\end{tabular}
\caption{Ratios of $\bar{B}^0$, $\bar{B}_s^{0}$, $\Lambda_b^0$ and $\Xi_b^{0,-}$ to $B^-$ at $T_{H}$=170 and 160\,MeV in the PDG and RQM scenarios.}
\label{tab_rat}
\end{table}
%
%

{\it Bottom-hadron $p_T$-spectra in $pp$ collisions.---}
To compute the $p_T$ differential cross sections of ground-state $b$ hadrons, we simulate the fragmentation and decay processes
using the $b$-quark $p_t$ spectrum in $\sqrt{s}=5.02$, $7$ and $13$\,TeV $pp$ collisions from FONLL~\cite{Frixione:2007nw,Cacciari:2012ny}. A $b$ quark sampled
from the spectrum is fragmented into $b$ hadrons via the same FF~\cite{Kartvelishvili:1977pi} as implemented in FONLL,
\begin{equation}
D_{b\rightarrow H_b}(z)\propto z^{\alpha}(1-z) \ ,
\label{bFF}
\end{equation}
where $z=p_T/p_t$ is the fraction of the $b$-hadron's ($H_b$) momentum, $p_T$, over the $b$-quark momentum, $p_t$. The fragmentation weight
of $H_b$ is determined by its thermal density $n_{H_b}^{\rm primary}$, Eq.~(\ref{primary_density}), normalized by the sum $\sum_{H_b} n_{H_b}^{\rm primary}$.
Each $H_b$ produced from fragmentation is then decayed into the ground-state particles with a constant matrix element, \ie, decay kinematics solely
determined by phase space and $BR$s estimated above.

\begin{figure}[!t]
\hspace{-0.95cm}
\begin{minipage}{0.27\textwidth}
\includegraphics[width=1.2\textwidth]{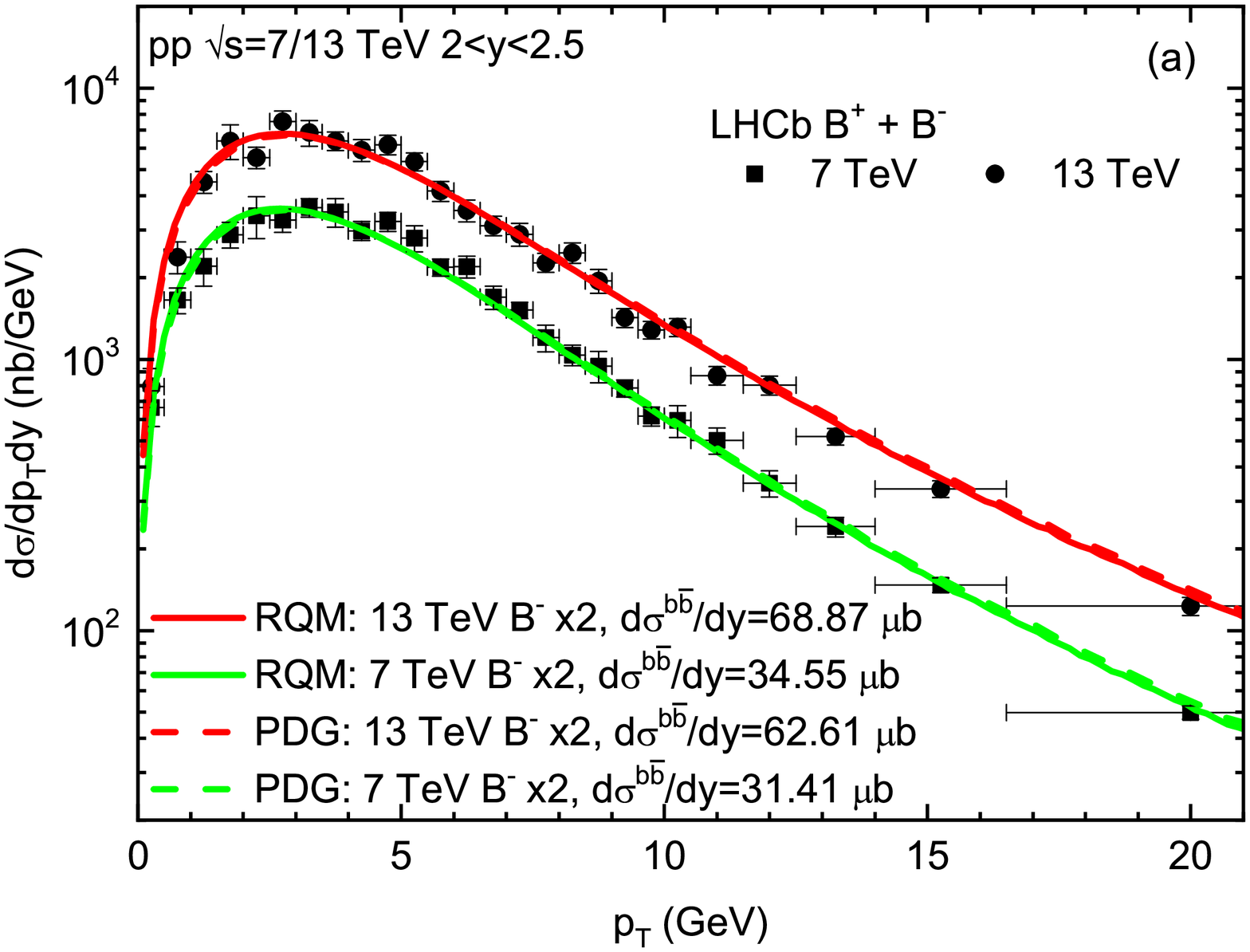}
\end{minipage}
\hspace{-0.5cm}
\begin{minipage}{0.27\textwidth}
\includegraphics[width=1.2\textwidth]{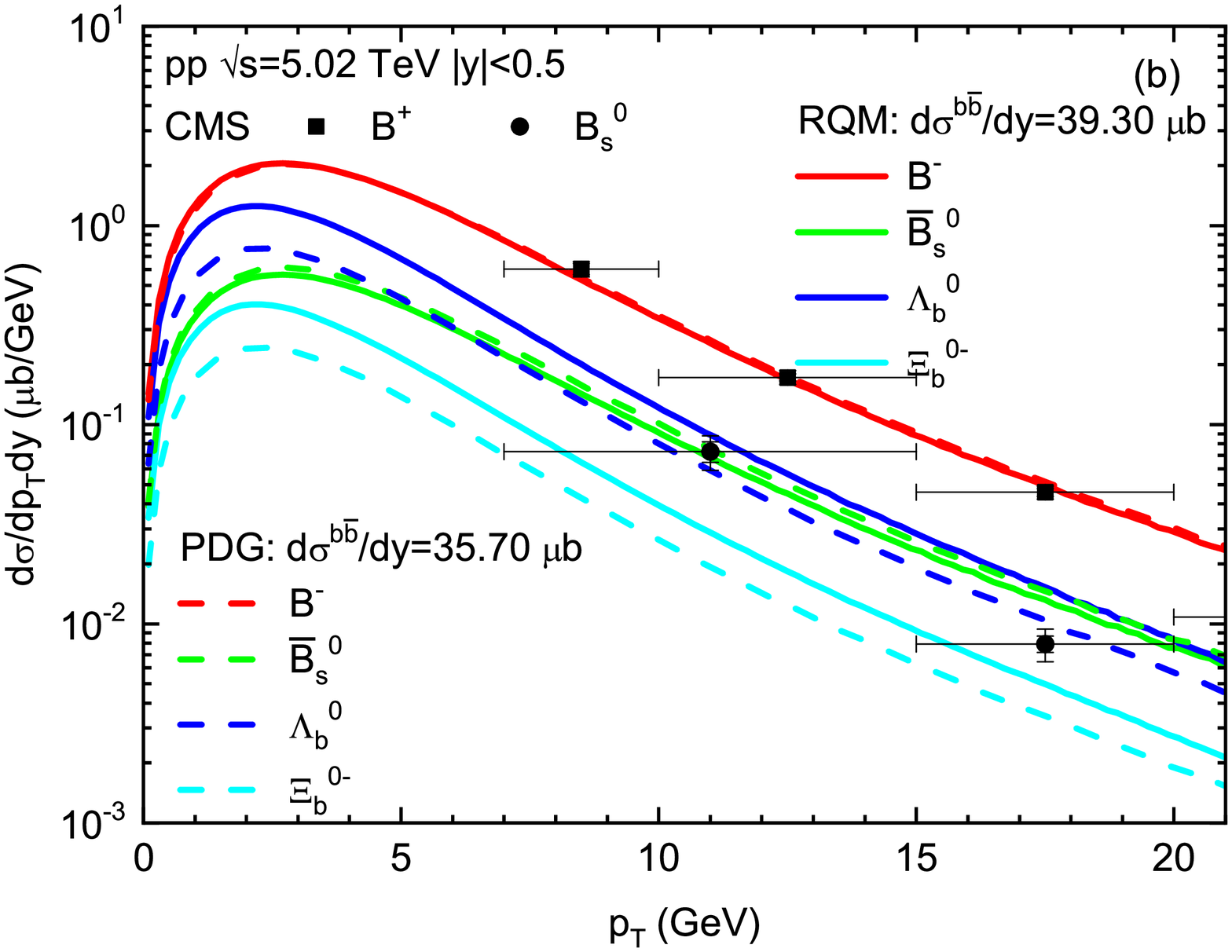}
\end{minipage}

\caption{(a): $p_T$-differential cross sections of ground-state $B^-$ in $\sqrt{s}$=7 and 13\,TeV $pp$ collisions for the RQM (solid lines) and PDG (dashed lines) scenarios at $T_H$=170\,MeV, in comparison with LHCb data at $2<y<2.5$~\cite{LHCb:2017vec}.
(b): the same for $B^-$, $\bar{B}_s^0$, $\Lambda_b^0$ and $\Xi_b^{0,-}$ in $\sqrt{s}$=5.02\,TeV $pp$ collisions at mid-rapidity, compared to $|y|<2.4$ CMS data~\cite{CMS:2017uoy,CMS:2018eso} scaled to $|y|<0.5$ via FONLL~\cite{Frixione:2007nw,Cacciari:2012ny}.}
\label{fig_pTspectra}
\end{figure}

\begin{figure*}[!thb]
\begin{minipage}[c]{0.3\textwidth}
\vspace{-0.5cm}
\hspace{-0.4cm}
\includegraphics[width=1.18\textwidth]{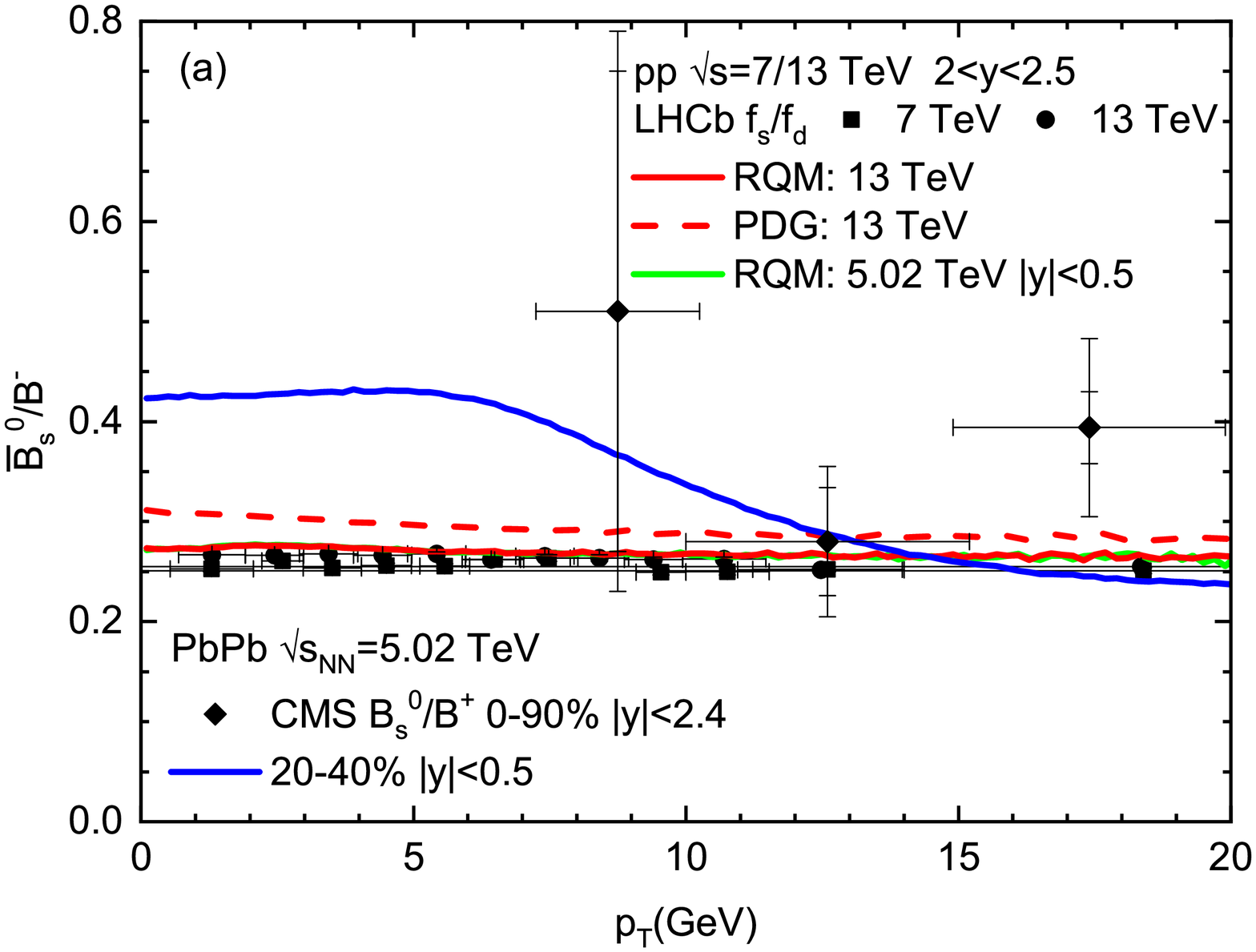}
\end{minipage}
\begin{minipage}[c]{0.3\textwidth}
\vspace{-0.5cm}
\hspace{-0.4cm}
\includegraphics[width=1.18\textwidth]{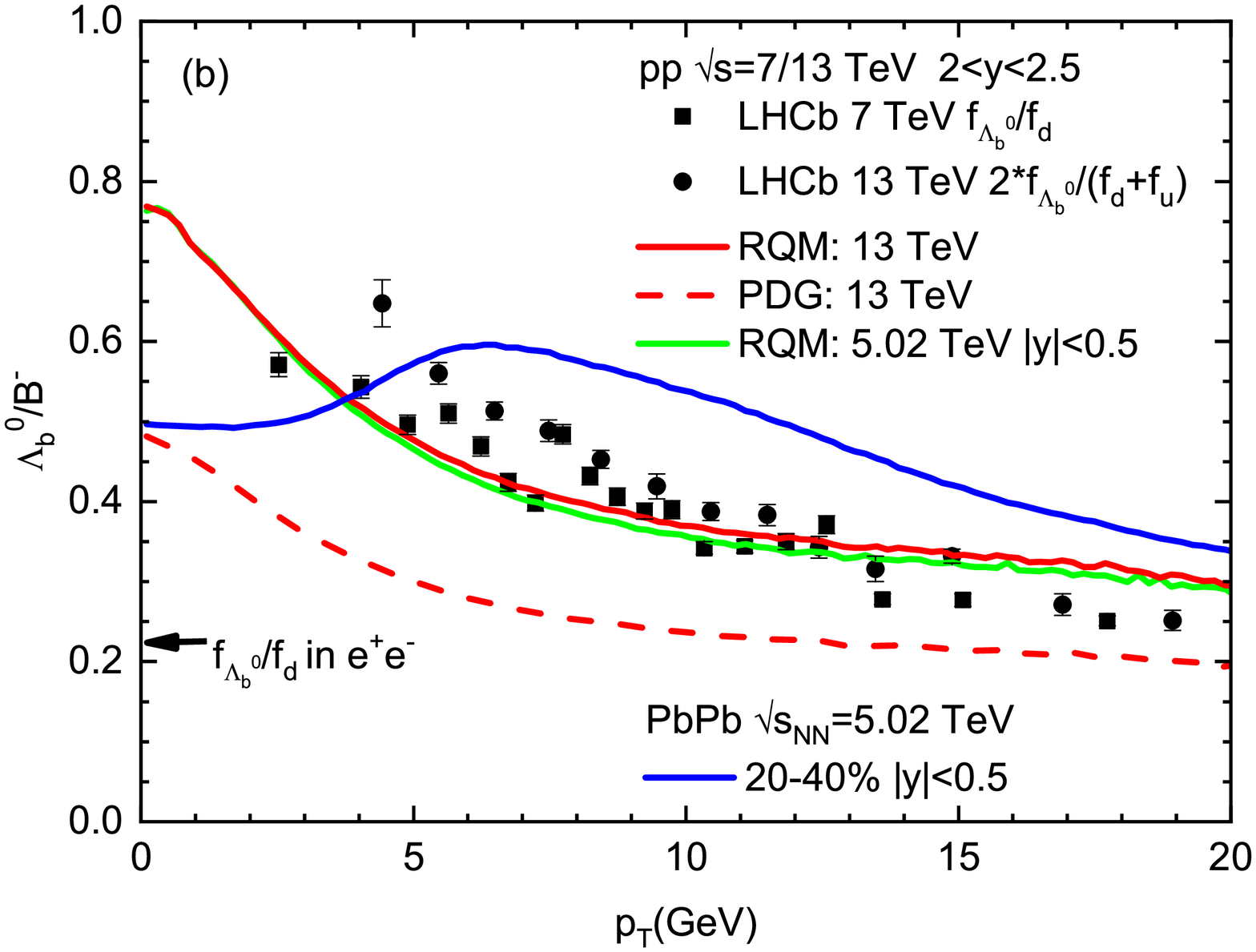}
\end{minipage}
\begin{minipage}[c]{0.3\textwidth}
\vspace{-0.5cm}
\hspace{-0.4cm}
\includegraphics[width=1.18\textwidth]{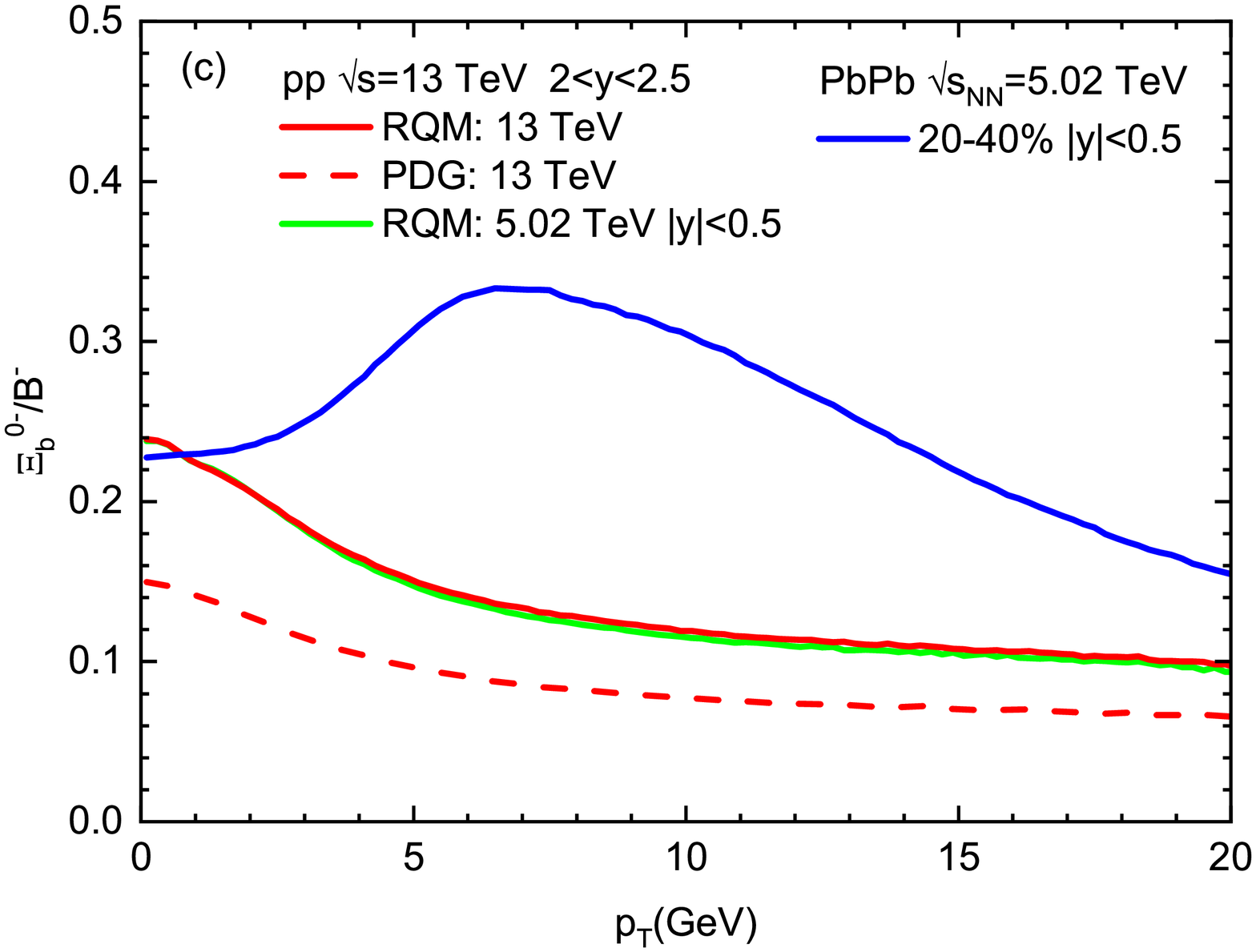}
\end{minipage}

\vspace{-0.3cm}

\caption{$p_T$-dependent ratios of (a): $\bar{B}_s^0/B^-$, (b): $\Lambda_b^0/B^-$ and (c): $\Xi_b^{0,-}/B^-$  for
RQM (solid lines) and PDG (dashed lines) scenarios in $\sqrt{s}$=13\,TeV (red, $2<y<2.5$) and 5.02\,TeV (green, $|y|<0.5$) $pp$ collisions, together with ratios
in $\sqrt{s_{NN}}$=5.02\,TeV PbPb collisions (blue solid lines, 20-40\% centrality) at mid-rapidity, in comparison with available LHCb~\cite{LHCb:2019lsv,LHCb:2014ofc,LHCb:2019fns} and CMS~\cite{CMS:2021mzx} data. The horizontal arrow in the middle panel indicates the LEP average for $f_{\Lambda_b}/f_d$~\cite{HFLAV:2019otj} from $Z^0$ decays at an average $b$-quark transverse momentum of $\langle p_t(b)\rangle$$\sim$40\,GeV.}
\label{fig_pTratios}
\end{figure*}

The parameter $\alpha$ in Eq.~(\ref{bFF}) is tuned to fit the slope of the $p_T$ spectra of ground state $b$ hadrons (kinematic effects from recombination are partially absorbed by this tune). For the RQM scenario at $T_H=170$\,MeV, we find that with $\alpha_B=45$ (for simplicity taken the same for all $B$ mesons; similarly $\alpha_{B_s}=25$ for all $B_s$ mesons and $\alpha_{\rm baryon}=8$ for all $b$ baryons), the measured $B^++B^-$ $p_T$-differential cross section at $2<y<2.5$~\cite{LHCb:2017vec} can be described with a total $b\bar{b}$ cross section of $d\sigma^{b\bar{b}}/dy=34.55~\mu b$ ($68.87~\mu b$) in $\sqrt{s}=7$\,TeV (13\,TeV) $pp$ collisions, cf.~Fig.~\ref{fig_pTspectra}(a). The latter is a prediction based on the
computed hadro-chemistry content and is consistent with the LHCb data for semileptonic $b$ decays~\cite{LHCb:2016qpe}. For the PDG
scenario, the $B^++B^-$ data turn out to be equally well described but with a $\sim10\%$ smaller total $b\bar{b}$ cross section. The decrease in the
latter is due to the reduction of $b$ content in the baryon sector, as demonstrated by the significantly smaller $p_T$-differential yields of
$\Lambda_b^0$ and $\Xi_b^{0,-}$ compared to their RQM counterparts, cf.~Fig.~\ref{fig_pTspectra}(b) for $\sqrt{s}=5.02$\,TeV $pp$ collisions at mid-rapidity,
where available data for $B^+$ and $B_s^0$ are compared; the value $d\sigma^{b\bar{b}}/dy=39.30~\mu b$ deduced from the RQM scenario is
comparable to the value measured by ALICE via non-prompt $D$ mesons~\cite{ALICE:2021mgk}.

We now turn to the $p_T$-dependent $b$-hadron ratios. For $\bar{B}_s^0/B^-$, shown in Fig.~\ref{fig_pTratios}(a), the additional states in the RQM
scenario reduce the PDG scenario results by over 10\% at low $p_T$, improving the description of the LHCb data for $f_s/f_d$ which are
approximately constant with $p_T$.
For $\Lambda_b^0/B^-$, the RQM scenario is clearly favored by the LHCb data, cf.~Fig.~\ref{fig_pTratios}(b). The substantial gap between data
and the PDG scenario results is largely overcome by the feeddown of the large set of ``missing" baryons included in the RQM calculation, leading to a fair
description of the data, including its increasing trend toward low $p_T$. For comparison, the LEP average of
$f_{\Lambda_b}/f_d$~\cite{HFLAV:2019otj} in $Z$ decays is indicated as a horizontal arrow. Finally, our predictions for the  $\Xi_b^{0,-}/B^-$ ratio from
RQM and PDG scenarios are compared in Fig.~\ref{fig_pTratios}(c), exhibiting similar features as in the case of $\Lambda_b^0/B^-$.

%
%
{\it Bottom hadrons in PbPb collisions.---}
The hadro-chemistry computed above in $pp$ collisions serves as a controlled reference for studying modifications in heavy-ion collisions.
Toward this end, we employ a strongly coupled transport approach previously developed for the $c$ sector~\cite{He:2019vgs} and calculate
the hadro-chemistry and nuclear modification factor of $b$ hadrons in $\sqrt{s_{NN}}$=5.02\,TeV PbPb collisions. In this approach, the $b$ quark
diffusion in the hydrodynamically evolving QGP is simulated via relativistic Langevin equations whose accuracy is improved compared to $c$ quarks
because of the $\sim 3$ times larger $b$-quark mass (no shadowing is put on the initial $b$-quark spectrum).
The transport coefficient is taken from lQCD-potential based $T$-matrix computations~\cite{Riek:2010fk} but amplified by the
same $K$=1.6 factor as done for the $c$ sector~\cite{He:2019vgs}, to mimic missing contributions from spin-dependent
forces~\cite{ZhanduoTang:2023tdg} and/or radiative energy loss.
At $T_H$=170\,MeV $b$-quark hadronization into mesons/baryons is computed by the 4-momentum conserving resonance recombination model (RRM)~\cite{Ravagli:2007xx,He:2019vgs}.
The RRM is implemented event-by-event in combination with the Langevin diffusion based on selfconsistently determined recombination probabilities,
$P_{i}(p_b^*)$~\cite{He:2019vgs}. The sum of $\sum_i P_{i}(p_b^*)$ over all primary $b$ hadrons $i$ is renormalized to unity at vanishing $b$ quark restframe momentum ($p_b^*$) to guarantee the majority of low-momentum $b$ quarks hadronize through recombination, while leftover $b$ quarks, as a result of the decreasing $\sum_i P_{i}(p_b^*)$ toward large $p_b^*$, fragment in the same manner as in $pp$. A prominent feature of our RRM implementation is
the inclusion of space-momentum correlations (SMCs) in the quark phase-space distributions~\cite{He:2019vgs}, which augments the flow effect in the
$p_T$-spectra of high-mass $b$ hadrons and generally extends the reach of recombination toward higher $p_T$.

In practice, constituent-quark ($m_{u,d}$=0.33\,GeV, $m_s$=0.45\,GeV, $m_b$=4.88\,GeV) and -diquark masses (scalar $m_{[ud]}$=0.71\,GeV within $\Lambda_b$'s, axial-vector $m_{\{ud\}}$=0.909\,GeV within $\Sigma_b$'s, $m_{[us]}$=0.948\,GeV and $m_{\{us\}}$=1.069\,GeV within $\Xi_b$'s, and $m_{\{ss\}}$=1.203\,GeV within $\Omega_b$'s) are taken from RQM studies~\cite{Ebert:2009ua,Ebert:2011kk}. We use energy-dependent widths (cf.~Ref.~\cite{Vovchenko:2018fmh}) with onshell values of  $\Gamma^0\sim0.1$\,GeV
in the meson, diquark and baryon cross sections in RRM, which suppresses artificial low-$s$ tails of the pertinent Breit-Wigner amplitudes for
$\sqrt{s}$ values far below the nominal resonance mass (thereby significantly reducing the sensitivity of final results to variations of the width values).

\begin{figure}[!t]
\hspace{-0.95cm}
\begin{minipage}{0.27\textwidth}
\includegraphics[width=1.2\textwidth]{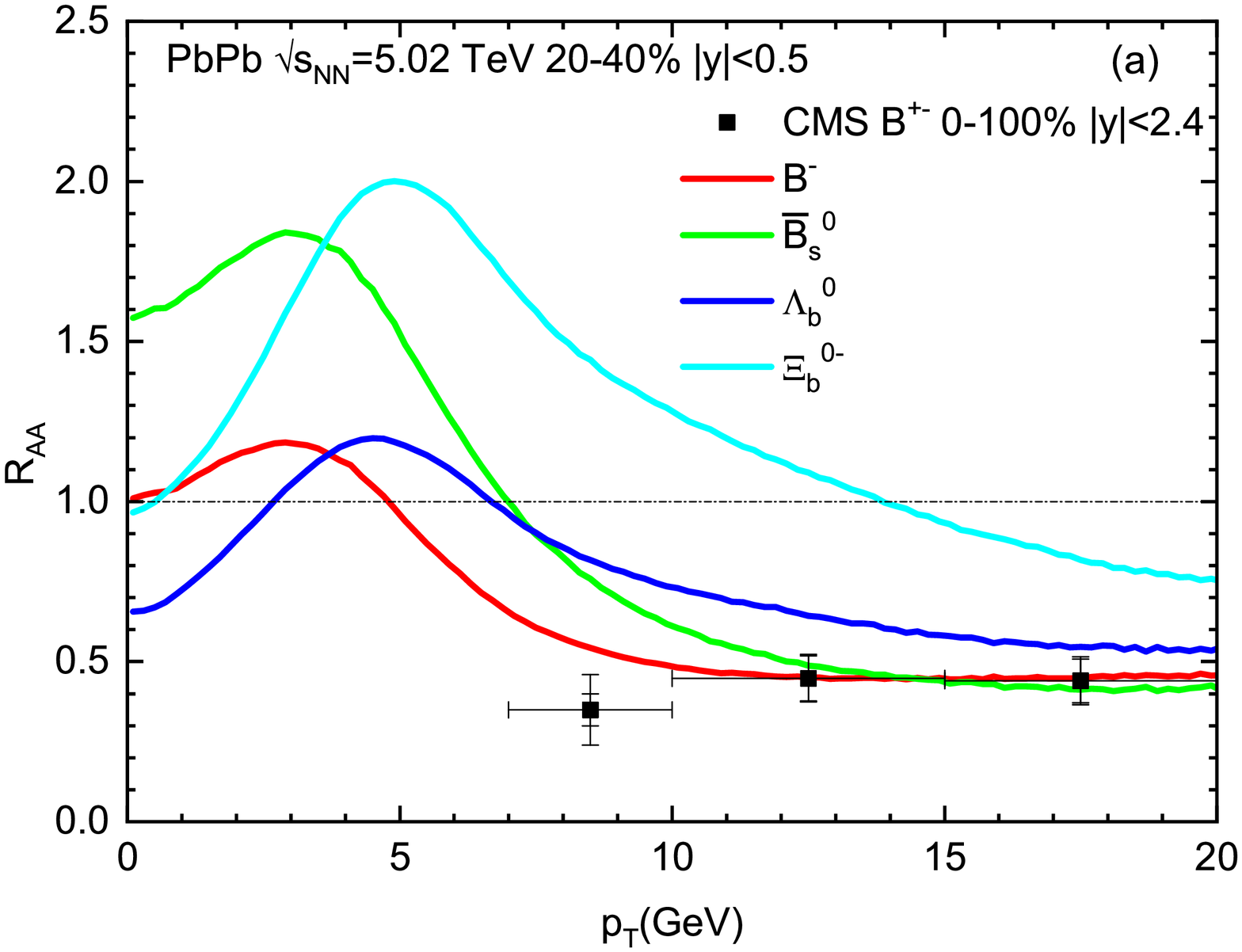}
\end{minipage}
\hspace{-0.5cm}
\begin{minipage}{0.27\textwidth}
\includegraphics[width=1.2\textwidth]{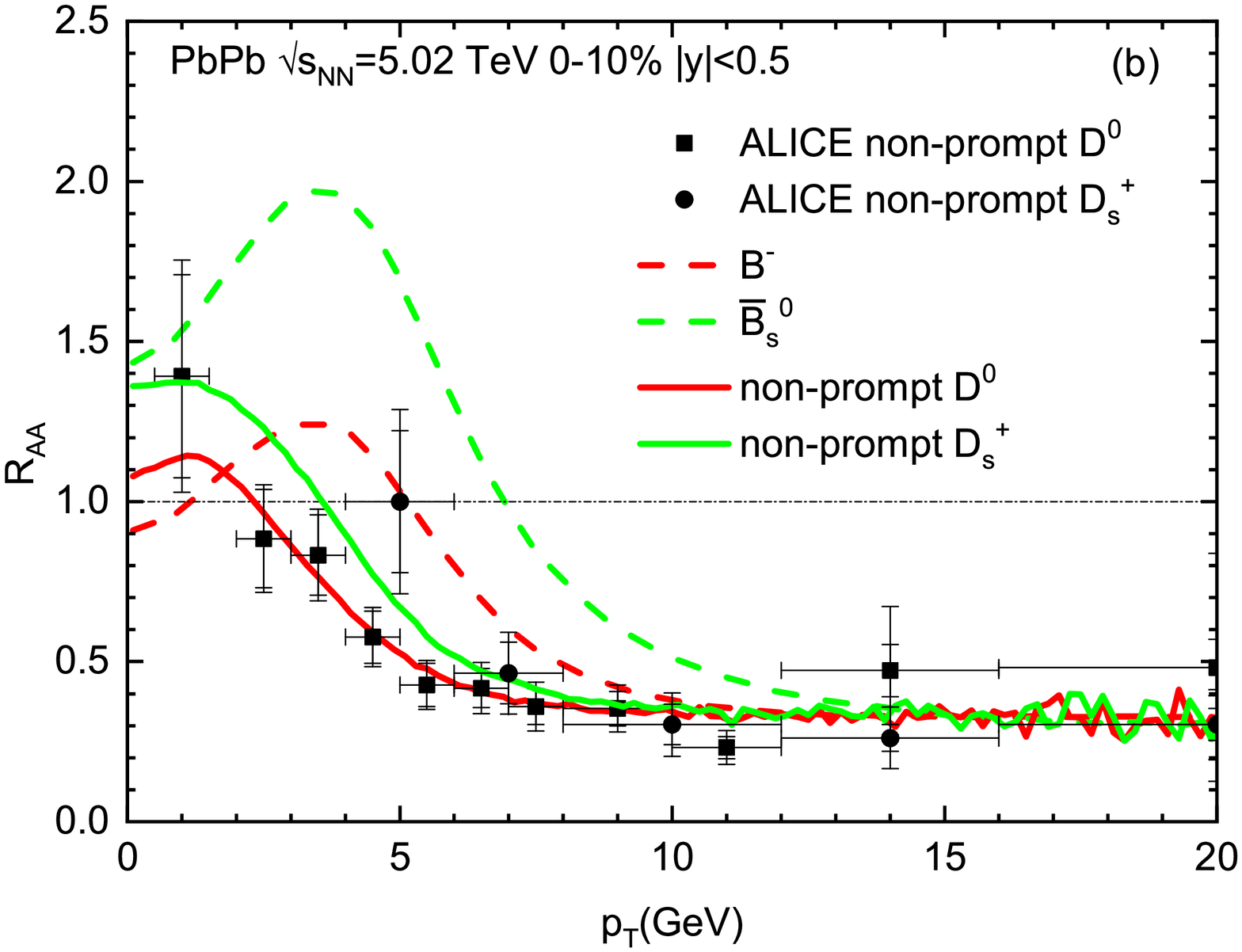}
\end{minipage}
\caption{(a): nuclear modification factors for ground-state $B^-$, $\bar{B}_s^0$, $\Lambda_b^0$ and $\Xi_b^{0,-}$ in 20-40\% $\sqrt{s_{NN}}$=5.02\,TeV PbPb collisions at mid-rapidity, together with CMS data for $B^{+-}$ in 0-100\% centrality~\cite{CMS:2017uoy}.
(b): the same for non-prompt $D^0$ and $D_s^+$ (0-10\% centrality) in comparison with ALICE data~\cite{ALICE:2022tji,ALICE:2022xrg}.}
\label{fig_RAA}
\end{figure}

Primary $b$ hadrons formed from hadronization undergo further diffusion in the hadronic phase until kinetic freezeout using our previously calculated
$D$-meson thermalization rate~\cite{He:2011yi} scaled down by the $b$-hadron mass. These hadrons are then decayed to obtain the $p_T$ spectra of
ground-state $b$ hadrons. The ratios of $\bar{B}_s^0/B^-$, $\Lambda_b^0/B^-$ and $\Xi_b^{0,-}/B^-$ are shown in Fig.~\ref{fig_pTratios}. Compared to their counterparts in $pp$ collisions, the $\bar{B}_s^0/B^-$ ratio exhibits a significant enhancement up to $p_T\sim$10\,GeV, resulting from $b$-quark coupling to the enhanced strangeness in QGP through recombination; an enhanced $\Lambda_b^0/B^-$ ratio appears in the intermediate-$p_T$ region due to a
stronger flow effect on generally heavier baryons as captured by RRM with SMCs, peaking at a higher $p_T\sim$6\,GeV and extending to significantly larger $p_T\sim$15\,GeV than the corresponding ratio in the $c$ sector~\cite{He:2019vgs,ALICE:2021bib} because of the larger $b$ quark mass.
The $\Xi_b^{0,-}/B^-$ ratio develops a more pronounced enhancement as it combines the strange-quark and baryon features. The nuclear modification
factors, $R_{AA}$, defined as the ratio of $p_T$ differential yield in PbPb collisions to the cross section in $pp$ collisions scaled by the nuclear overlap
function~\cite{ALICE:2018tvk}, are shown in Fig.~\ref{fig_RAA}(a) for ground-state $B^-$ (same for $\bar{B}^0$), $\bar{B}_s^0$, $\Lambda_b^0$ and
$\Xi_b^{0,-}$ in semicentral PbPb collisions, with an expected hierarchy of flow effects and suppression driven by their quark content.
Upon weak-decay of these hadrons into non-prompt $c$ hadrons utilizing Pythia8~\cite{Sjostrand:2014zea}, the resulting $R_{AA}$'s for non-prompt $D^0$ and $D_s^+$ in central collisions show fair agreement with ALICE data, cf.~Fig.~\ref{fig_RAA}(b).

{\it Summary.---}
Employing the statistical hadronization model we have evaluated the hadro-chemistry of $b$ hadrons in $pp$  collisions at collider energies. The spectrum
of $b$-hadrons has been taken from theoretical predictions of the relativistic quark model which is largely supported by lattice-QCD
computations of vacuum spectroscopy. Many of the RQM states, especially in the baryon sector, are not yet observed and are therefore much more
numerous than the current PDG listings. With strong and electromagnetic feedown estimated from the $^3P_0$ model, we have performed quark
model estimates of excited $b$-hadron decays which enabled a comprehensive prediction of fragmentation fractions of weakly-decaying $b$ hadrons for the
first time; pertinent ratios turn out to agree with measurements in $p\bar{p}$ collisions at the Tevatron. We have further calculated $p_T$ differential
cross sections for ground-state $b$ hadrons using fragmentation weights of an underlying (perturbative) $b$-quark spectrum determined by the SHM. The
resulting $p_T$-dependent $\bar{B}_s^0/B^-$ and $\Lambda_b^0/B^-$ ratios agree with LHCb data. All of our results critically
depend on the excited states that the RQM predicts beyond the PDG listings.
We have furthermore deployed the new hadro-chemsitry into our strongly-coupled HF transport model for heavy-ion collisions, to evaulate spectral
modifications in PbPb collisions. The $p_T$-dependent modifications of ratios between different ground-state $b$ hadrons have been quantified, highlighting
the role of $b$ quarks as probes of the QGP -- a central pillar of experimental efforts at both RHIC~\cite{Dean:2021rlo} and the LHC in the near future~\cite{Citron:2018lsq,ALICE3:leterofintent2022,LHCb:2022ine,ALICE:2022wpn}.

{\it Acknowledgments.--}
This work was supported by NSFC grant 12075122 and the U.S.~NSF under grant no.~PHY-1913286. MH thanks Fabrizio Grosa for help with the simulation of weak decays of $b$ hadrons.

\end{document}